\documentclass[10pt,twocolumn,english,aps,prl,twocolum,superscriptaddress,bibnotes,amsmath,amssymb,floatfix]{revtex4-1}
\usepackage[colorlinks=true,citecolor=blue,linkcolor=magenta]{hyperref}
\usepackage{soul}
\usepackage[utf8]{inputenc}
\usepackage[english]{babel}
\usepackage{amsmath,amsfonts,amssymb}
\usepackage[T1]{fontenc}
\usepackage{url}
\usepackage{graphicx}
\graphicspath{{./Figures/}}

\begin{document}

\title{Arbitrary-phase locking of fiber Mach-Zehnder interferometers}

\author{Ruiyang Chen}
\affiliation{International Quantum Academy, Shenzhen 518048, China}
\affiliation{Shenzhen Institute for Quantum Science and Engineering, Southern University of Science and Technology, Shenzhen 518055, China}

\author{Yi-Han Luo}
\email{luoyh@iqasz.cn}
\affiliation{International Quantum Academy, Shenzhen 518048, China}

\author{Jinbao Long}
\affiliation{International Quantum Academy, Shenzhen 518048, China}

\author{Junqiu Liu}
\affiliation{International Quantum Academy, Shenzhen 518048, China}
\affiliation{Hefei National Laboratory, University of Science and Technology of China, Hefei 230088, China}


\begin{abstract}
 
\end{abstract}

\maketitle

\noindent\textbf{Optical interferometers are extensively used in fundamental physics test, gravitational wave detection, quantum metrology, topological photonics, and quantum information processing. 
Fiber-based interferometers are compact, robust and cheap, thus are ubiquitously deployed. 
However, the optical phase in fiber interferometers is sensitive to ambient perturbation, resulting in compromised phase sensing precision. 
Therefore, phase control, shifting and stabilization of fiber interferometers is essential.
Methods to create stable interference patterns and to lock a fiber interferometer at arbitrary phase have been shown, which however are sophisticated, bulky and delicate, preventing wider application in harsh environment outside laboratories or in space. 
Here we demonstrate a new method for arbitrary-phase locking of fiber unbalanced Mach-Zehnder interferometers. 
Compared to existing method, our method is simpler, more robust and more compact.  
We showcase the preparation and characterization of narrow-band energy-time-entanglement photon state generated in integrated nonlinear microresonators, where two-photon interference visibility reaching $0.993(6)$ is enabled. 
Our method constitutes a critical building block for photonic quantum network, and is useful to emerging single-photon interference in curved space-time that facilitates exploration of the interface of quantum mechanics and general relativity.}

Optical phase is extensively used for coding information in modern telecommunication systems. 
As is sensitive to environment perturbation, optical phase also provides a route for sensing and measurement, e.g. to precisely characterize temperature, pressure and vibration along the fiber path. 
Unlike optical intensity, direct detection of optical phase is challenging. 
Thus optical interferometers, mediating optical phase and intensity, are developed and employed. 
Today optical interferometers are equally essential in emerging applications including fundamental physics test \cite{Vallone:16, WuHN:24}, 
gravitational wave detection \cite{Amaro-Seoane:17, Armano:18}, 
quantum metrology \cite{Nagata:07, Afek:10}, 
topological photonics \cite{Regensburger:12, ChenC:18, ChenC:23}, 
and quantum information processing \cite{Marcikic:04, Gisin:07, Farrera:18, Tchebotareva:19, YuY:20, Zhong:20, Madsen:22}.
In these applications, stable interference is critical. 
For example, the quantum interference of single photon travelling through curved space-time can be exploited to identity the interface of quantum mechanics and general relativity \cite{Zych:11, XuP:19, Terno:20}. 
To detect  gravitational red-shift-introduced tiny phase difference $\delta \phi$, the interferometer's phase fluctuation must be suppressed orders of magnitude lower than $|\delta\phi|$.
In quantum communication, the fidelity of quantum information encoding and decoding critically rely on the interferometers' stability at the transmitter and receiver ends. 

\begin{figure}[t!]
\centering
\includegraphics[width=0.45\textwidth]{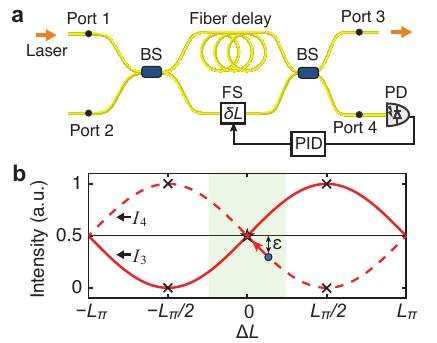}
\caption{
\textbf{Principle of a fiber unbalanced Mach-Zehnder interferometer (UMZI)}.
\textbf{a}. 
The common phase-locking method of a fiber UMZI. 
FS, fiber stretcher to add a fiber length $\delta L$.
\textbf{b}. 
The UMZI's output optical intensity $I_3$ (from Port 3) and $I_4$ (from Port 4) as a function of $\Delta L$
For example, locking $I_3=0.5$ (red star) requires locking $I_4=0.5$ via reducing the error signal $\epsilon$. 
There are two locking regions, the green-shaded region where $I_4$ has negative slope and negative feedback is enabled; 
and the region where $I_4$ has positive or near-zero slope and the negative feedback fails. 
}
\label{Fig:1}
\end{figure}

Fiber interferometers featuring compactness, robustness and low cost are commonly used. 
However, contrast to interferometers based on free-space optics, the optical interference in fiber interferometers can be severely corrupted by thermal and mechanical disturbances. 
Consequently, active feedback control and stabilization of optical phase are critical in these interferometers and their applications. 
Unbalanced Mach-Zehnder interferometers (UMZI) are the most widely used fiber interferometers. 
The phase control and stabilization in a UMZI is illustrated in Fig.~\ref{Fig:1}a.  
A laser of frequency $f_0$ enters Port 1, while Port 2 is idle. 
The laser is evenly splitted into two branches by a beam splitter (BS). 
The fiber length in the upper branch is longer than that of the lower branch by $L$.
The two branches recombine on another BS and create optical interference.
The phase difference $\Phi$ between the two branches linearly depends on $L$, as $\Phi=2\pi nLf_0/c=\pi L/L_\pi$, 
where $n$ is the refractive index of optical fibers, 
$c$ is the speed of light in vacuum,
and $L_\pi=c/(2nf_0)$ is the unit length for $\pi$-phase shift.
Neglecting optical loss in optical fibers and BS, we apply $L=L_0+\Delta L$, where $L_0=2NL_\pi~(N\in\mathbb{N}_+)$ and $|\Delta L| < L_\pi \ll L_0$.
Thus
\begin{equation}
    \Phi=2\pi N+\pi\Delta L/L_\pi, 
    \label{eqn:phase_origin}
\end{equation}
The normalized output optical intensity at Port 3 varies as $I_3=(1+\sin\Phi)/2=[1+\sin (\pi \Delta L/L_\pi)]/2$, as shown in Fig. \ref{Fig:1}b solid curve. 

Experimentally, locking $\Phi$ is equivalent to stabilize $\Delta L$. 
However, due to ambient thermal and mechanical perturbation, $\Delta L$ fluctuates temporally as $\Delta L(t)$. 
Commonly, a feedback loop is required to stabilize $\Delta L(t)$.
Figure~\ref{Fig:1}b dashed line shows the normalized output optical intensity $I_4=[1-\sin (\pi \Delta L/L_\pi)]/2$ at Port 4 probed by a photodetector (PD).
Due to the unitary nature of BS \cite{zeilinger:81}, $I_4$ has a $\pi$-phase shift to $I_3$.
For example, the stabilization of $\Delta L=0$ requires locking $I_4$ to the set-point $I_0=0.5$, marked as the star in Fig. \ref{Fig:1}b. 
To do so, the electric signal from the PD is fed into a PID module, which generates a control signal based on the error signal $\epsilon=I_4(\Delta L)-I_0$. 
The control signal drives the fiber stretcher (FS), which varies the fiber length of the lower branch, such that $\epsilon=0$ is maintained, resulting in $\Delta L=0$. 

However, the above method has two limitations. 
First, the PID's output control signal enables negative feedback if and only if $I_4(\Delta L)$ is monotonic.  
Exemplified in Fig. \ref{Fig:1}b, if the PID allows negative feedback on the negative slope of $I_4(\Delta L)$ (green-shaded region), the feedback becomes positive on the positive slope of $I_4(\Delta L)$ and thus invalid for locking. 
Second, for locking near $\Delta L\sim \pm L_\pi/2$ marked with crosses in Fig. \ref{Fig:1}b, $I_4(\Delta L)$ is insensitive to $\Delta L$ variation, nullifying phase locking. 
Consequently, once the PID is configured, the locking range is limited, e.g. the green-shaded region in Fig. \ref{Fig:1}b. 

\begin{figure*}[t!]
\centering
\includegraphics[width=1\textwidth]{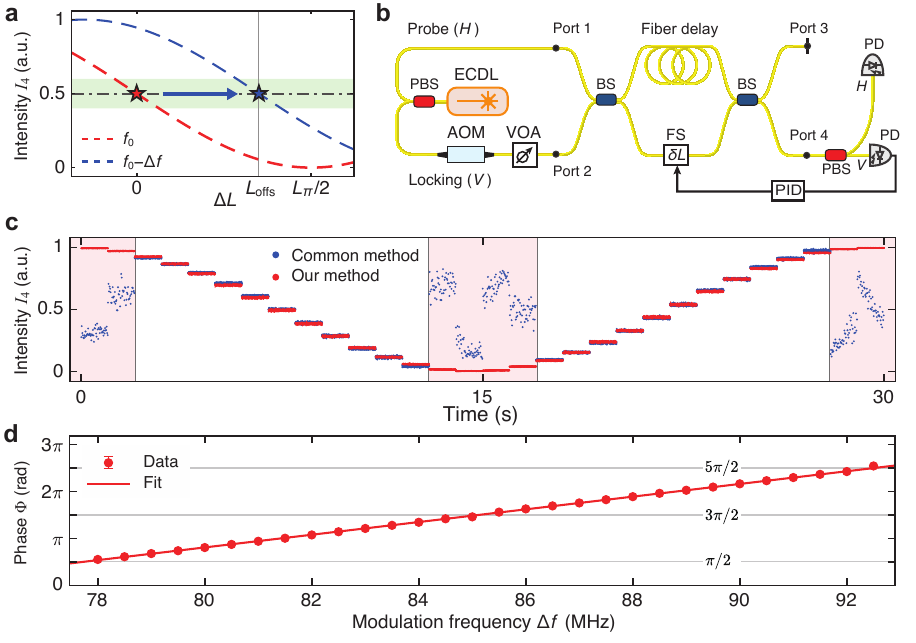}
\caption{
\textbf{Illustration of our frequency-shifted method for arbitrary-phase locking of a fiber UMZI}.
\textbf{a}. 
Principle of our phase-locking method. 
To lock $\Delta L$ at any value, the locking laser is frequency-shifted by $-\Delta f$, such that $I_4$($\Delta L$) curve is translated by $L_\mathrm{offs}$, i.e. from the red curve to the blue curve. 
Stabilizing $I_4$ at $I_0=0.5$ leads to locking $\Delta L=L_\mathrm{offs}$, corresponding to the translation from the red star to the blue star.
The slope at the red star is identical to that at the blue star, thus the feedback remains equally effective. 
\textbf{b}. 
Experimental setup.
\textbf{c}. 
Comparison of phase locking performance. 
As the AOM's modulation frequency $\Delta f$ is increased stepwise, $I_{4,H}$ is measured over 1 second for each stepped $\Delta f$. 
Our method (red dots) allows locking $I_{4,H}$ to any value, while the common method (blue dots) fails in the region near $I_{4,H}=0$ and $1$ (red-shaded region).
\textbf{d}.
Extracted phase $\Phi$ as a function of $\Delta f$, and the linear fit. 
Error bar is plotted as the standard deviation of $\Phi$ at each step, which is however much smaller than the dot size and thus is invisible. 
}
\label{Fig:2}
\end{figure*}

To overcome these issues, several strategies have been implemented \cite{Freschi:95, Rogers:16, Roztocki:21}, where a phase-locking laser consisting of multiple frequency components is used. 
In Ref. \cite{Freschi:95, Rogers:16}, the phase-locking laser is sinusoidally phase-modulated to generate sidebands. 
In Ref. \cite{Roztocki:21}, a portion of the phase-locking laser is frequency-shifted by an acousto-optic modulator (AOM). 
The misaligned interference fringes corresponding to different frequency components are measured, digitized, and processed with micro-controllers to calculate the real-time phase.
This phase is compared to the set-point, and a digital control signal is generated. 
The digital signal is then converted to an analogue signal driving the fiber stretcher. 
In such a way, the phase is varied and stabilized. 

\begin{figure*}[t!]
\centering
\includegraphics[width=1\textwidth]{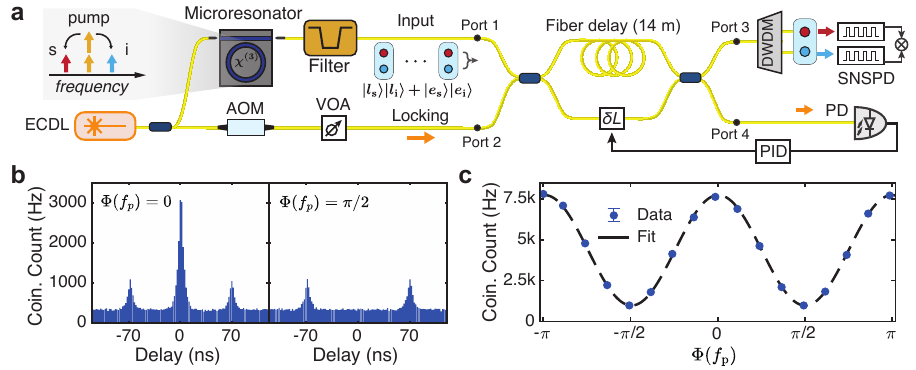}
\caption{
\textbf{Characterization of narrow-band, energy-time entangled photon pairs using our phase-locking method}
\textbf{a}. 
Experimental setup. 
\textbf{b}.
Two-photon correlation histogram. 
In the left/right panel, the peak at zero delay appears/vanishes due to constructive/destructive two-photon interference. 
\textbf{c}.
Two-photon interference fringe. 
The fit indicates a raw visibility of $V=0.777(8)$. 
By subtracting the background, the visibility is improved to 0.993(6).
Error bar is plotted, however much smaller than the dot size and thus invisible. 
}
\label{Fig:3}
\end{figure*}

Here we demonstrate an innovative phase-locking method that is much simpler, more robust and efficient compared with existing methods. 
The principle is following. 
To stabilize $\Delta L$ at any value, instead of changing the set-point from $I_0=0.5$, we shift the laser frequency $f_0$ to $f_0-\Delta f$. 
Thus Eq. \ref{eqn:phase_origin} becomes
\begin{align}
 \notag\Phi(f_0-\Delta f) & = 2\pi n (L_0+\Delta L)(f_0-\Delta f)/c \\[2pt]
 \notag & = 2\pi n (L_0+\Delta L-L_\mathrm{offs})f_0/c  \\[2pt]
\label{eqn:phase} & = 2\pi N + \pi (\Delta L -L_\mathrm{offs}) / L_\pi
\end{align}
where $\Delta f \ll f_0$, 
$L_\mathrm{offs}=L_0\Delta f/f_0$, 
and the secondary term $\Delta f_0 \Delta L$ is neglected. 
A comparison of Eq. \ref{eqn:phase} with Eq. \ref{eqn:phase_origin} suggests that, the frequency shift of $-\Delta f$ translates $I_4(\Delta L)$ by $L_\mathrm{offs}$. 
This is illustrated as the translation from the red curve to the blue curve in Fig. \ref{Fig:2}a.
Thus now, for the laser of frequency $f_0-\Delta f$, stabilizing $I_4$ at $I_0=0.5$ leads to locking $\Delta L=L_\mathrm{offs}$, as the translation from the red star to the blue star in Fig. \ref{Fig:2}a. 
Note that, the blue curve's slope at the blue star ($\Delta L=L_\mathrm{offs}$) is identical to the red curve's slope at the red star ($\Delta L=0$), thus the feedback remains equally effective. 
Meanwhile, sufficiently large $L_0$ enables $L_\mathrm{offs}=L_0\Delta f/f_0>2L_\pi$. 
Therefore, with proper $\Delta f$, $\Delta L$ can be locked to any value within $[-L_\pi, L_\pi]$, as
\begin{gather}
    \Delta L=\frac{L_0}{f_0}\Delta f
    \label{eqn:dldf}
\end{gather}

Figure \ref{Fig:2}b shows our experimental setup to quantify the phase-locking performance.
The output from an external-cavity diode laser (ECDL) is splitted into two branches. 
In the probe branch, light is oriented to horizontal (H) polarization and enters Port 1 of the UMZI. 
In the locking branch, light is oriented to vertical (V) polarization, frequency-shifted by an AOM, and enters Port 2. 
Meanwhile, a variable optical attenuator (VOA) is used to compensate the AOM's transmission variation with different modulation frequency $\Delta f$.
The UMZI has $L_0\approx14$ m ($\sim69$ ns time delay), and is placed in a heat-insulated container to suppress high-frequency phase fluctuation. 
Upon exiting Port 4, the light components for probe and locking are separated by a polarizing beam splitter (PBS) and probed by two PDs.

Experimentally, the AOM's modulation frequency $\Delta f$ is increased stepwise from 78.0 MHz to 92.5 MHz, resulting in linearly increasing $L_\mathrm{offs}$. 
Thus stabilizing the output optical intensity of V polarization at Port 4, i.e. $I_{4,V}=I_0=0.5$ and $\Phi(f_0-\Delta f)=2\pi N$, leads to locking $\Delta L=L_\mathrm{offs}$. 
With the linearly increasing $L_\mathrm{offs}$ and thus $\Delta L$, the detected $I_{4,H}$ of H polarization varies sinusoidally as $\Phi(f_0)=2\pi N+\pi\Delta L/L_\pi$. 
In sum, we have
\begin{equation}
\left\{
\begin{aligned}
    I_{4,V}&=0.5\\
    I_{4,H}&=\frac{1+\sin (\frac{\pi L_0}{L_\pi f_0} \Delta f)}{2}\\
\end{aligned}
\right.
\label{Eq.IVH}
\end{equation}

For each stepped value of $\Delta f$, Figure \ref{Fig:2}c red dots show the measured $I_{4,H}$ over 1 second. 
Meanwhile, the locking performance of $I_{4,H}$ using our method is compared with that using the common method (blue dots). 
The latter is implemented by removing the AOM and VOA. 
Note that for negative and positive slope of $I_{4,H}$ (white regions), different PID configurations are optimized. 
It is clear that, while the common method fails in the region near $I_{4,H}=0$ and $1$ (red-shaded region), corresponding to $\Phi(f_0)=\pi/2,3\pi/2$ and $5\pi/2$, our method remains equally effective at any value.
Each stepper $\Phi$ is extracted from measured $I_{4,H}$ using Eq. \ref{Eq.IVH}, whose average value and standard deviation are calculated. 
Figure \ref{Fig:2}d shows the extracted $\Phi$ as a function of $\Delta f$, and the linear fit.  
By averaging the standard deviation for each stepped $\Phi$, we estimate the phase-locking precision to be 0.011 rad.
This corresponds to stabilizing $\Delta L$ with precision of $\lambda_0/561$, where $\lambda_0=c/f_0$ is the optical wavelength in vacuum.
For example, we have $\lambda_0/561=2.76$ nm for $\lambda_0=1550$ nm. 

To showcase an application of our phase-locking method, we characterize two-photon interference visibility of narrow-band, energy-time entangled photon pairs \cite{Franson:89}, which are critical quantum light sources for long-distance quantum communications \cite{Marcikic:04, Gisin:07, Farrera:18, Tchebotareva:19, YuY:20}.  
Figure \ref{Fig:3}a shows the experimental setup. 
The ECDL's output is divided into two branches.
The lower branch is again the locking branch. 
In the upper branch, the laser of frequency $f_\mathrm{p}$ is coupled into an integrated silicon nitride (Si$_3$N$_4$) microresonator of intrinsic quality factor exceeding $10^7$ \cite{Ye:23, Chen:24} and 100 GHz free spectral range (FSR). 
Via cavity-enhanced spontaneous four-wave mixing (SFWM) \cite{Helt:10, Luo:15} in the Si$_3$N$_4$ microresonator, two photons in the pump laser annihilate, creating a pair of signal and idler photons aligned to the microresonator's resonance grid. 
The signal and idler photons have $f_\mathrm{s}$ and $f_\mathrm{i}$ frequency.
Energy conservation requires $2f_\mathrm{p}=f_\mathrm{s} + f_\mathrm{i}$.

Due to random generation time of photon pairs, the photon pair $|e_\mathrm{s}\rangle|e_\mathrm{i}\rangle$ created earlier is superposed with that created later $|l_\mathrm{s}\rangle|l_\mathrm{s}\rangle$, resulting in an entangled state $|\Psi\rangle=(|e_\mathrm{s}\rangle|e_\mathrm{s}\rangle+|l_\mathrm{s}\rangle|l_\mathrm{s}\rangle)/\sqrt{2}$, where s/i denotes signal/idler photon. 
To measure two-photon interference fringe, the pump laser is filtered out and a UMZI is used to overlay photons generated at different time \cite{Franson:89}.  
The signal and idler photons are separated using a dense wavelength-division multiplexer (DWDM) after the UMZI, and detected with superconducting nanowire single-photon detectors (SNSPD). 

With the experimental setup, we project the signal/idler photon along the state $|\phi_\mathrm{s/i}\rangle=(|e_\mathrm{s/i}\rangle+e^{i\Phi(f_\mathrm{s/i})}|l_\mathrm{s/i}\rangle)/\sqrt{2}$, where $\Phi(f_\mathrm{s/i})$ is the UMZI's phase for the signal/idler photon. 
The probability of measuring $|\Psi\rangle$ to be $|\phi_\mathrm{s}\rangle|\phi_\mathrm{i}\rangle$ is 
\begin{equation}
\begin{aligned}
    p&=\frac{1}{4} + \frac{1}{4}\cos\left[\Phi(f_\mathrm{s})+ \Phi(f_\mathrm{i})\right]\\
    &=\frac{1}{4} + \frac{1}{4}\cos 2\Phi(f_\mathrm{p})\\
\end{aligned}
\label{Eq.5}
\end{equation}
Here we use $2\Phi(f_\mathrm{p})=\Phi(f_\mathrm{s})+ \Phi(f_\mathrm{i})$ due to $2f_\mathrm{p}=f_\mathrm{s} + f_\mathrm{i}$ and $\Phi(f)=2\pi n Lf/c$. 
Experiment, we vary the AOM's modulation frequency $\Delta f$ to stabilize $\Phi(f_\mathrm{p})$ within the range $[-\pi, \pi]$.

The the photon arrival events detected by SNSPD are recorded and analyzed with a time tagger.
The two-photon correlation histogram, describing the two-photon arrival time difference, is shown in Fig. \ref{Fig:3}b, where two-photon interference is evidenced.
When $\Phi(f_\mathrm{p})=0$, the central peak reaches the maximum and is fourfold to the sidebands. 
When $\Phi(f_\mathrm{p})=\pm\pi/2$, the central peak vanishes due to destructive interference. 

To obtain interference fringe and extract interference visibility, the zero-delay peak in Fig. \ref{Fig:3}b is post-selected. 
The coincidence count rate $n_\mathrm{cc}$ is calculated by summing up the bins within the temporal range $[-2.8, 2.8]$ ns. 
Figure \ref{Fig:3}c shows the measured $n_\mathrm{cc}$ versus $\Phi(f_\mathrm{p})$, where $\Phi(f_\mathrm{p})$ is calculated from $\Delta f$ using Eq. \ref{eqn:phase_origin} and \ref{eqn:dldf}. 
The raw interference visibility $V$ is extracted by fitting $n_\mathrm{cc}$ with $n_\mathrm{cc}=0.5A[1+V\cos\Phi(f_\mathrm{p})]$, where $V=0.777(8)$ at $P=689~\mu$W pump power and $A$ is another fit parameter. 
As shown in Fig. \ref{Fig:3}b, the background far from zero delay is substantial due to spontaneous Raman scattering \cite{Karpov:16}, which heavily deteriorates interference visibility. 
We calculate the background by averaging the bin values far from the zero delay, and subtract the background for each data point in Fig. \ref{Fig:3}c. Finally a visibility of 0.993(6) is achieved.

In conclusion, we demonstrate a simple and efficient method allowing arbitrary-phase locking for fiber UMZI with a fully analog PID feedback. 
The phase-locking system requires neither digital signal processing, nor analog-digital conversion. 
In addition, upon shifting the PID set-point, its slope remains maximized, resulting in equally efficient PID performance.  
For a UMZI of arm length difference of $~14$ meter, we experimentally lock the phase difference $\Phi$ to any value within $[0, 2\pi]$ and with 0.011 rad precision. 
The precision can be further improved by using a high-speed fiber stretcher enabling larger feedback bandwidth.

We use the stabilized UMZI to showcase the preparation and characterization of two-photon interference visibility of integrated, narrow-band, energy-time entangled photon pairs. 
Our method enables two-photon interference visibility of 0.993(6), evidencing the qualification to encode and decode information in quantum communication systems. 
It also benefits emerging single-photon interference experiments in curved space-time, which can facilitates exploration of the interface of quantum mechanics and general relativity. 
Furthermore, using ultralow-loss integrated waveguides \cite{Liu:21, Puckett:21, Zhang:17}, integrated AOM \cite{Stanfield:19, Tian:20, Liu:20a} or EOM \cite{WangC:18, HeM:19}, and VOA \cite{Rickman:14}, our method can be translated to integrated photonics, allowing photonic-chip-based interferometers for integrated quantum systems and networks.

\medskip
\begin{footnotesize}

\noindent \textbf{Acknowledgments}: 
We thank Yuan Cao and Hui-Nan Wu for the fruitful discussion.
J. Liu acknowledges support from the National Natural Science Foundation of China (Grant No.12261131503), Innovation Program for Quantum Science and Technology (2023ZD0301500), Shenzhen-Hong Kong Cooperation Zone for Technology and Innovation (HZQB-KCZYB2020050), and Shenzhen Science and Technology Program (Grant No. RCJC20231211090042078).
The silicon nitride chips were fabricated by Qaleido Photonics. 

\noindent \textbf{Author contributions}: 
Y.-H. L. conceived the experiment. 
R. C. and Y.-H. L. built the experimental setup and performed the experiment, assisted with J. Long. 
R. C., Y.-H. L. and J. Liu analyzed the data and wrote the manuscript. 
J. Liu supervised the project.  

\noindent \textbf{Conflict of interest}:
The authors declare no conflicts of interest. 

\noindent \textbf{Data Availability Statement}: 
The code and data used to produce the plots within this work will be released on the repository \texttt{Zenodo} upon publication of this preprint.

\end{footnotesize}


%

\end{document}